\documentclass[10pt,twocolumn,twoside]{IEEEtran}

\usepackage[dvipsnames]{xcolor}
\usepackage{comment}
\usepackage{tikz}
\usetikzlibrary{external}
\usetikzlibrary{mindmap,trees}
\usetikzlibrary{arrows}
\usetikzlibrary{calc}
\usepackage{adjustbox}
\usepackage{subcaption}

\usepackage{amsthm}

\newtheorem{lemma}{Lemma}
\newtheorem{remark}{Remark}
\newtheorem{definition}{Definition}

\usepackage{amsmath,amsfonts,amssymb,color}
\DeclareMathOperator*{\argmin}{argmin}
\usepackage{epsfig}
\usepackage{psfrag}
\usepackage{algorithm}
\usepackage{algpseudocode}
\usepackage{array}
\usepackage{physics}
\usepackage{epstopdf}
\usepackage{tikz, cite}
\usepackage{cleveref}

\allowdisplaybreaks

\begin{document}

\title{\LARGE \bf Combining Hybrid Input-Output Automaton and Game Theory for Security Modeling of Cyber-Physical Systems}
\author{Mustafa Abdallah,~Sayan Mitra,~Shreyas Sundaram, and~Saurabh Bagchi
\thanks{Mustafa Abdallah, Shreyas Sundaram, and Saurabh Bagchi are with the School of Electrical and Computer Engineering at Purdue University. Email: {\tt \{abdalla0,sundara2,sbagchi\}@purdue.edu}. Sayan Mitra is with  the School of Electrical and Computer Engineering at the University of Illinois at Urbana-Champaign.  Email: \tt mitras@illinois.edu.
}
}

\maketitle
\thispagestyle{empty}
\pagestyle{empty}

\begin{abstract}
We consider a security setting in which the Cyber-Physical System (CPS) is composed of subnetworks where each subnetwork is under  ownership of one defender. Such CPS can be represented by an attack graph where the defenders are required to invest (subject to a budget constraint) on the graph's edges in order to protect their critical assets (where each defender's critical asset has a certain value to the defender if compromised). We model such CPS using Hybrid Input-Output Automaton (HIOA) where each subnetwork is represented by a HIOA module. We first establish the building blocks needed in our setting. We then present our model that characterizes the continuous time evolution of the investments and discrete transitions between different system's states (where each state represents different condition and/or perturbation) within the system. Finally, we provide a real-world CPS example to validate our modeling.
\end{abstract}

\IEEEpeerreviewmaketitle
\section{Introduction}

Cyber-physical systems (CPS) are often composed of hardware and software blocks that are not highly reliable in themselves. However, some CPS applications demand a high degree of criticality, i.e., safety, security, and reliability~\cite{humayed2017cyber,abdallah2020morshed}. Such CPS applications are increasingly facing attacks by sophisticated adversaries which motivates the fundamental problem we set out to solve --- how to create such CPS applications out of the inherently unreliable building blocks. In this context, a significant line of research has been performed on understanding how to better secure CPS \cite{laszka2015survey,sanjab2017prospect,9030279}. This research involved several mathematical frameworks that have been developed for modeling cyber-physical systems, each with their attendant strengths and weaknesses.

There exists several challenges, that were not tackled in the literature, for modeling CPS precisely~\cite{garlan2015modeling}. At a very high level, a model should describe the state of the system and how that state can change. For example, the state variable of the model of an autonomous vehicle has to include variables representing physical quantities like position, velocity, and angular speed of wheels, etc., as well as variables representing the state of the software modules used for perception, planning, and control~\cite{chen2017cyber}. Moreover, a CPS model, broadly speaking, has two kinds of such variables: continuous variables, such as security mechanisms that can be modeled as continuous variables (e.g., the fraction of traffic that is monitored for malicious packets on a network link) and discrete variables (e.g., the number of security personnel to deploy to a given site). Therefore, the question that need to be answered is how to model a CPS that involves both such kinds of variables precisely.

In addition to defining the state variables, a CPS model also has to describe how the values of these variables can change. Such changes are naturally described by programs and the natural language for describing the laws of the physical world is the language of ordinary differential equations (ODE). Bringing together discrete-continuous variables, programs, and ODEs within the same mathematical model gives rise to the so-called hybrid models. In this regard, several different model classes fall under the umbrella term hybrid systems, such as hybrid automata \cite{lygeros2003dynamical}, hybrid input/output automata~\cite{kaynar2010theory,mitra2007verification}, hybrid dynamical systems~\cite{goedel2012hybrid}, and switched systems~\cite{liberzon2003switching}. 

In all of these works, there exist two fundamental gaps between the goal of modeling resilient CPS precisely and the current state-of-the-art, in the areas of modeling, security, game theory, and distributed algorithms for CPS. The two gaps laid out next, taken together, prevent us from building highly resilient CPS applications. First, the models typically do not capture all the facets required to answer the two modeling requirements (e.g., they may focus on detailed element-level modeling or only the static modeling that can inform only the deployment decision). Second, the security algorithms typically only consider elements in isolation and when they do consider interdependent systems, the algorithms are oblivious to the requirements that arise due to the legacy nature of assets or the presence of multiple stakeholders, e.g., the fact that not all assets can be secured and that there may be limits on interactions among the stakeholders. 

Exceptions include the recent works~\cite{abdallah2020behavioral,woods2020network,hota2018game} that studied the interdependency between multiple stakeholders with a security game setting, and provided a method to calculate the optimal investments by the defenders to minimize their loss. However, they did not model the continuous time nature of the system and the transitions between different system states. In other words, they only solved (partially) the second gap in the state-of-the-art.

In this paper, we combine Hybrid Input-Output automaton modeling with game theory --- to the best of our knowledge, this hybrid has never been attempted before. We demonstrate that this hybrid can be put to good use to model CPS involving multiple defenders who are responsible for defending interdependent subnetworks within the system. Fundamentally, our hybrid modeling enables us to model both continuous and discrete transitions of large-scale CPS together. Specifically, we build-up our modeling framework based on the hybrid I/O automata (HIOA) of ~\cite{kaynar2010theory,mitra2007verification}. We choose this framework because it explicitly identifies input/output variables and actions of the automata, which makes it particularly suitable for defining externally visible interfaces across different types of modules (or players) in the CPS in a precise manner. 

\noindent {\bf Relevance to Big Data Workshop}. An implicit, but crucial factor that can enable our modeling is the availability of enormous amounts of data. Such data would be needed to fit the state changes of the discrete as well as continuous variables. The fidelity of our modeling approach and consequently its utility depends on such ``big data" being collected and then synthesized to generate the model parameters. A second tie-in is the application domain that seems a natural fit to our hybrid modeling formulation --- embedded systems that are often the core of autonomous systems like robotics and autonomous driving. Being able to secure such systems will be a big victory for big data techniques. Therefore we feel that in this line of work, big data has a crucial role to play in improving the security modeling of autonomous systems. 

The remainder of this paper is organized as follows. We introduce the preliminaries and notations of our framework in Section \ref{sec: notations}, followed by the proposed HIOA framework in Section \ref{sec: HIOA_classes}. In Section \ref{sec: Example}, we apply our framework to a real-world CPS. We conclude the paper in Section~\ref{sec: conclusion}. 
\section{Preliminaries and Notations}\label{sec: notations}
In this section, we introduce the notations that we use later in our modeling framework, including the hybrid input-output automaton (HIOA) framework, and the general game-theoretic setup of the defenders of the CPS.

\subsection{Hybrid Input-Output Automaton (HIOA)}
A hybrid automaton is a useful model of a system that displays continuous-time behavior interleaved with discrete jumps. Hybrid automata with inputs and outputs have additional structure as they allow exogenous time-varying inputs, and observable outputs. Due to this additional structure, HIOA facilitate modular descriptions of subsystems and their compositions to obtain larger systems (as will be shown later in the paper).

A hybrid input/output automaton (HIOA) $A$ is defined as a tuple
			$$
			 (\mathcal{L}, \mathcal{X}, \mathcal{U}, \mathcal{M}, \mathcal{G}, \mathcal{R}, \Delta, \mathcal{T}, \mathcal{Y},
			 \mathcal{I})
			$$
\begin{itemize}
    \item $\mathcal{L}$ is a finite set of discrete modes.
    
    \item $\mathcal{X} = \{x_1, x_2, \dots, x_n\}$ is a finite set of $n$ state variables, and following the notation defined before, $X$ denotes the set of all valuations of $\mathcal{X}$. We denote any particular vector of states by $\mathbf{x} = (x_1, x_2,\dots, x_n)$. Thus, the hybrid state space is a subset of the set $\mathcal{L}\times X$.
    
    \item $\mathcal{U} = \{u_1, u_2, \dots, u_m\}$ denotes the set of $m$ typed input variables. Note that these variables can be of different types (e.g., Real ($\mathbb{R}$), Integers ($\mathbb{Z}$), or  Boolean). We denote $\mathbf{u}$ as the vector of the input variables where  $\mathbf{u} = (u_1, u_2, \dots, u_m)$.
    
    \item $\mathcal{M}$ maps each mode $l \in \mathcal{L}$ with a mode invariant $M(l) \in X \times\mathcal{U}$.
    
    \item $\mathcal{G}$ is a set of predicates over $X\times\mathcal{U}$.
    
    \item $\mathcal{R}$ is a set of functions from $X \times\mathcal{U}$ to $X$. 
    
    \item $\Delta \in \mathcal{L}\times\mathcal{G}\times\mathcal{R}\times\mathcal{L}$ is a finite set of transitions. For each transition $\delta \in \Delta$, $g \in \mathcal{G}$ is its guard predicate, and $r \in \mathcal{R}$ is its reset map.
    
    \item $\mathbb{T} \in \mathbb{R}_{\geq0}$ represent the domain of time values.
    
    \item A trajectory $\tau(\mathcal{X},\mathcal{U})$ is a function from $\mathbb{T}$ to $(X\times\mathcal{U})$ that describes the valuations of the input variables and state variables over time. Note that in an HIOA, a trajectory is often a sequence of alternating flows (within modes) and resets (consistent with mode transitions).
    
    \item  The set of all trajectories for the set of variables $V$ is denoted by $trajs(V)$ where $\mathcal{T} \subseteq trajs(V)$.
    
    \item $\mathcal{Y} \in \mathcal{X}$ denotes the set of typed output variables. 

    \item $\mathcal{I} \in \mathcal{L} \times X$ is the set of possible initial discrete modes and valuations of the state variables.
\end{itemize}
Note that for a given mode $l$ the flow within the mode is typically the solution trajectory $\mathbf{x}(\cdot)$ of an initial value problem as described by ODE $\mathbf{\dot{x}} = f_l(\mathbf{x},\mathbf{u})$ with the initial condition $v(\mathbf{x}) = \mathbf{x}_0$ at 
time $t = t_0$.

In addition, $A$ satisfies the following axioms:

$\mathbf{E_1}$ (Input transition enabled) For every $l \in \mathcal{L}$ and $a \in \Delta$, there exists $l' \in \mathcal{L}$ such that $l
\xrightarrow[]{a} l'$.

$\mathbf{E_2}$ (Input trajectory enabled) For every $l \in \mathcal{L}$ and every $v \in trajs(\mathcal{U})$, there exists $\tau \in \mathcal{T}$, such that $\tau.fstate = l$, $\tau \downarrow \mathcal{U} \leq v$, and either (a) $\tau \downarrow \mathcal{U} = v$, or (b) $\tau$ is closed and
some $l \in \mathcal{L}$ is enabled in $\tau.lstate$.\footnote{Note that $\tau.fstate$ and $\tau.lstate$ are the first state and last state of a trajectory $\tau$, respectively.  Also, $a \downarrow b$ denotes the restriction of the function $a$ into the set $b$~\cite{mitra2007verification}.}

\begin{remark}
The axioms of input transition enabling $(\mathbf{E_1})$ and the input trajectory enabling $(\mathbf{E_2})$ are necessary for composition properties (that will be discussed later) to hold.
\end{remark}

\subsection{Properties of HIOA}
For any large CPS system, it makes sense to start with smaller modules and then put the modules together to create increasingly larger and more complex pieces until we build the whole system. For instance, operators of large-scale CPS have subordinates operating subsystems of this CPS. The benefits of that
modular approach are that repeated modules can be reused, and concurrency can be exploited. In this context, we exploit such powerful properties of HIOA to represent large-scale CPS. In this context, we introduce one main property of HIOA that are useful in our setting of modeling CPS.
\subsubsection{Closure under decomposition}
We build large HIOA models from smaller modules, using the \textit{composition} operation which is denoted by $||$. Composing two HIOA $A_1$ and $A_2$ results in a new object $A = A_1||A_2$. The theory of HIOA defines
compatibility conditions on the components that ensure that $A$ is also a valid HIOA. This property is called \textit{closure under composition}. For example, consider a cyber attack scenario involving a networked CPS such as
the Power grid in Section~\ref{sec: Example}, where each different subnetwork is managed by a different defender. In the HIOA framework, each of these components would be represented as an automaton. 

\subsection{Game Theoretic Framework}

In this subsection, we describe our general security game framework, including the attack graph representation of a CPS and the characteristics of defenders and attackers.

\subsubsection{Attack Graph}
We represent the assets in a CPS as nodes of a directed graph $ G=(V,\mathcal{E}) $ where each node $v_{i} \in V $ represents an asset. A directed edge $(v_{i},v_{j}) \in \mathcal{E}$ means that if node $v_{i}$ is successfully attacked, it can be used to launch an attack on node $v_{j}$.  We assume that the success of attacks across different edges in the network are captured by independent random variables. Each edge $(v_i, v_j) \in \mathcal{E}$ has an associated weight $p_{i,j}^0 \in (0,1]$, denoting the probability of successfully attacking asset $v_j$ starting at $v_i$ (in the absence of any security investments).
The graph contains a designated source node $ v_{s} $, which is used by the attacker to begin her attack on the network. For a general asset $ v_{t} \in V$, we define $\mathcal{P}_{t}$ to be the set of directed paths from the source $ v_{s} $ to $v_{t}$ on the graph, where a path $ P \in \mathcal{P}_{t}$ is a collection of edges $ \lbrace{(v_{s}, v_{1}), (v_{1}, v_{2}), . . . , (v_{k}, v_{t})\rbrace} $.
  Therefore, in the absence of any security investments, the probability that $v_{t}$ is compromised due to an attacker exploiting a given path $ P \in \mathcal{P}_{t}$ is $ {\displaystyle \prod_{(v_{m},v_{n}) \in P} p_{m,n}^0} $, by our aforementioned independence assumption. The attacker can choose any path from the multiple attack paths in $ \mathcal{P}_{t} $ to attack $v_t$. Figure~\ref{fig:nescor-attack-graph} shows an example of attack graph modeling of CPS (here, the failure of smart power grid).

\subsubsection{Strategic Defenders}
Let $ \mathcal{D} $ be the set of all defenders of the network. Each defender $ D_{k} \in \mathcal{D} $ is responsible for defending a subnetwork (i.e., a set $ V_{k} \subseteq V \setminus \lbrace{v_s\rbrace} $ of assets). For each compromised asset $ v_{m} \in V_{k} $, the defender $ D_{k} $ will incur a financial loss $L_{m} \in \mathbb{R_{\geq \texttt{0}}} $. To reduce the attack success probabilities on edges interconnecting assets inside the network, a defender can allocate security resources on these edges, subject to the constraints described below.

Let $\mathcal{E}_k \subseteq \mathcal{E}$ be the subset of edges that defender $D_k$ can allocate security resources on.  We assume that each defender $D_k$ has a security budget $B_k \in \mathbb{R}_{\ge 0}$.  Thus, we define the defense strategy space of each defender $D_k \in \mathcal{D}$ by
\begin{equation}
X_{k} \triangleq \lbrace{x_{i,j}^{k} \in \mathbb{R_{\geq \texttt{0}}} , (v_{i}, v_{j}) \in \mathcal{E}_{k} \!: \!\!\sum_{(v_{i}, v_{j}) \in \mathcal{E}_{k}} \!\!x_{i,j}^{k} \leq B_{k}  \rbrace}.
\label{eq:defense_strategy_space}
\end{equation}
In words, the defense strategy space for defender $D_k$ consists of all nonnegative investments on edges under her control, with the sum of all investments not exceeding the budget $B_k$. We denote any particular vector of investments by defender $D_k$ by $x_k \in X_k$.

Under a joint defense strategy, the total investment on edge $(v_i,v_j)$ is  
$x_{i,j}:=\{\sum_{D_k\in \mathcal{D}} x^k_{i,j}: (v_{i}, v_{j}) \in \mathcal{E}_{k}\}$. Let $p_{i,j}:\mathbb{R}_{\geq 0}\rightarrow [0,1]$ be a function mapping the total investment $x_{i,j}$ to an attack success probability, and with $p_{i,j}(0) = p_{i,j}^0$. 

The goal of each defender $ D_{k} $ is to choose her investment vector $x_k$ in order to best protect her assets from being attacked.  In this paper, we consider the scenario where each defender minimizes the highest probability path to each of her assets; this captures settings where the specific path taken by the attacker is not known to the defender a priori, and thus the defender seeks to make the most vulnerable path to each of her assets as secure as possible. Mathematically, this is captured via the cost function 
\begin{equation}\label{eq:defender_utility}
C_{k}(\mathbf{x}) = \sum_{v_{m} \in V_{k}} L_{m} \hspace{0.3mm} \Big( \hspace{0.3mm} \underset{P \in P_{m}}{\text{max}}\prod_{(v_{i},v_{j}) \in P} p_{i,j}({x_{i,j}}) \hspace{0.3mm} \Big)
\end{equation}
subject to $x_{k} \in X_{k}$. Note that $ C_{k}(\mathbf{x}) $ is a function of the investments of all defenders, and thus we denote the cost by $ C_{k}(x_{k},\mathbf{x}_{-k}) $ where $ \mathbf{x}_{-k} $ is the vector of investments by defenders other than $ D_{k} $.  Each defender chooses her investment vector $x_k \in X_k$ to minimize the cost $C_k(x_k, \mathbf{x}_{-k})$, given the investments $\mathbf{x}_{-k}$ by the other defenders.

\begin{remark}
For each HIOA module, we will consider the investments of other defenders as part of the inputs to that HIOA module. Thus, within specific modes, the valuation of the internal state variables will be calculated via the notion of  \it{best response} of that defender to the other defender's investments which we define below.
\end{remark}

\begin{definition}\label{def:BR}
The best response of player $D_k$ at a given investment profile $\mathbf{x}_{-k}$ by other defenders is the set $\mathbf{x}_k^* \triangleq \argmin_{\mathbf{x}_{k} \in X_k} C_k(\mathbf{x}_k, \mathbf{x}_{-k})$.
\end{definition}

The recent works~\cite{abdallah2020behavioral,hota2018game,abdallah2019impacts} studies the above security game setting, and provides a method to calculate the optimal investments by the defenders with respect to the cost function \eqref{eq:defender_utility}. However, they did not model the continuous time nature of the system and the transitions between different modes (states).  In the next section, we will combine HIOA with this game theoretic framework to model large-scale CPS. To the best of our knowledge, our proposed model is the first effort to model both adversarial and stochastic choices for security analysis. 

\section{The Proposed HIOA framework for modelling interdependent system with multiple defenders}\label{sec: HIOA_classes}

Now, after introducing the notations of  multiple-defender setup and the HIOA framework. We now introduce our model to capture the modes and the continuous time evolution of variables (within each mode) where the interdependent system contains different subnetworks with one defender responsible for defending each subnetwork (as shown earlier in Section~\ref{sec: notations}). 

Developing this extension of the framework and applying it for security analysis of the target applications. To the best of our knowledge, our work is the first step in this direction by introducing the notion of rewards (or utility functions).

Now, we introduce the model's main components: the modes of operation, the variables, the trajectories, and valuations.

\subsection{Modes of Operation}
In our model, each HIOA has four modes
of operation, that we now detail
\begin{itemize}
    \item {\it Startup mode}: This mode represents the initial state of each subnetwork (defender).
    \item {\it Normal mode}: In this mode, each subnetwork should  be in a normal operation status where the defender is allocating the investments by best responding to optimal investments of other defenders. 
    \item {\it Alternate mode}: In this mode, the defender alternates her investments from the normal mode. This can happen due to any external event or when one of the other defenders change her security investments.
    \item {\it Fail mode}:  This mode represents one or more node failures (i.e., when one of the subnetwork components is successfully compromised). 
\end{itemize}

Note that each mode is reachable from some other modes when triggered by specific events. For example, the ``normal" mode is reachable from ``alternate" mode by external stability event. On the other hand, alternate mode is reachable from normal mode by external event such as sensing an attack of different type or when other HIOA (resp. its defender) deviates from her current investments.

We emphasize that in a real system, it is certainly possible for the system to encounter a failure in startup mode. We choose not to model this behavior for simplicity in modeling and analysis.

\subsection{Input, State, and Output Variables}

In our model, each subnetwork (managed by defender $D_k \in \mathcal{D}$) can be viewed as a HIOA module with the following inputs, outputs and internal states:

\begin{itemize}
    \item The set of state variables $\mathcal{X}$ is $\{\mathbf{x}_k, \tau, \mathbf{p}_k^0\}$, where $\mathbf{x}_k$ is the defender's defense investment vector over the edges and $\mathbf{p}_k^0$ is the vector of initial attack probabilities over the edges.
    
    \item The set of input variables $\mathcal{U}$ is \{Attack\_Risk, Fail\_Event, $\mathbf{x}_{-k}$\}, where Attack\_Risk is an indicator of the risk on the subnetwork and has a value of 0 if there is no attack incident and non-zero otherwise, Fail\_Event represents the triggering event of failure (or compromise) and $\mathbf{x}_{-k}$ is the investment of all defenders except defender $D_k$.
    
    \item The set of output variables is \{$\mathbf{p}_k$, $\mathbf{x}_{k}$\}.
\end{itemize}

\begin{remark}
Note that the estimation of model's parameters (e.g., the Attack\_Risk) can be inferred from the alerts provided by intrusion detection sensors deployed in various parts of the CPS's subnetworks. Such collection of data have been a challenging issue, however, there are several recent efficient algorithms for collecting this data for a large-scale CPS (e.g., smart agriculture~\cite{chatterjee2020context} and Cyber attacks~\cite{hadar2019big}). Thus, it becomes feasible for us to collect such parameters to build our game theoretic and HIOA  model.
\end{remark}

\subsection{Trajectories and Valuations}
\begin{figure}[t]
\centering
  \includegraphics[width=0.9\linewidth]{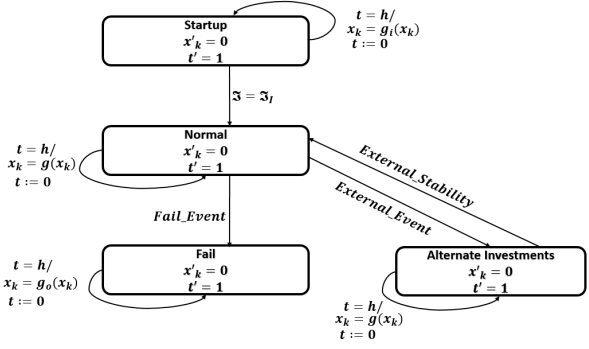}
  \caption{An Example HIOA module for one subnetwork. This HIOA module has four modes and the transitions between different modes are represented by directed arrows (where the corresponding conditions for such transitions are given above each arrow). The ODEs that represent the continuous time evolution of internal variables inside each mode are represented inside the four modes'. The valuations for the variables for each mode are also shown. Note that  $h$ denotes the sample period for the decision, where the continuous variable of time $t = m h$, and $m \in  \mathbb{R}_{\geq 0}$ is a sample number.}
  \label{fig:HIOA_Module}
\end{figure}

Now, we provide the trajectories that describe the relations between the different variables, the set of guards, and reset functions.
\begin{itemize}
    \item For any trajectory in $\mathcal{T}$, the flow function for the trajectory in any mode is described by the ODE $\dot{\mathbf{x}_k} = 0$.
    
    \item For each mode $l \in \mathcal{L}$, $\mathcal{M}$ maps $l$ to the negation of the conjunction of all the guards on its outgoing transitions. 

    \item The set of guards is \{Fail\_Event = true, $\tau = \tau_I\}$.
    
    \item The set of reset functions is a union of two functions $g(\cdot)$ and $g_o(\cdot)$. We present such functions in our update formulas.
    
    \item The transitions are as depicted in Figure~\ref{fig:HIOA_Module}. These transitions between different modes are represented by directed arrows (where the corresponding conditions for such transitions are given above each arrow). Note that the valuations functions are given in \eqref{eq: val_func_1}. Such valuation gives the probability of successful attack $\mathbf{p}_k(t)$, and the investments $\mathbf{x}_k(t)$ that minimize the cost in each mode (given by \eqref{eq:defender_utility}). The time horizon evolution is also shown.
    
    \item The set of initial states is the singleton set: \{(startup, $\mathbf{p}_k^0 \xrightarrow{} \mathbf{0}$, $\tau \xrightarrow{} 0$, $\mathbf{p}_k \xrightarrow{} \mathbf{0}$)\}. Noe that these set of initial states can be chosen for different CPS models based on the initial conditions of that CPS.
\end{itemize}
A summary of our HIOA module is given below.

\textbf{HIOA}: Subnetwork of defender $D_k$

\textbf{Variables}

\ \  \ \textbf{input:}
     Attack\_Risk: Boolean, investment of other defenders ($\mathbf{x}_{-k}$): Float.

\ \  \textbf{internal:}     
     Defense investments ($\mathbf{x}_{k}$): Real vector, Initial Success Prob. ($\mathbf{p}_{k}^0$): Real vector.
     
\  \  \textbf{output:}     
     Probability of Successful attack ($\mathbf{p}_k (t)$).
     
\textbf{Real trajectories}
\begin{flalign}\label{eq: val_func_1}
  \mathbf{p}_k (t) &= \mathbf{p}_{k}^0 \  f(\mathbf{x}_{k}(t))
\end{flalign}
  \[ \mathbf{x}_k(t) =    \left\{
    \begin{array}{ll}
          0 & \textit{if} \   Attack\_Risk = 0 \\
          min_{\mathbf{x}_{k} \in X_k} \  C_k(\mathbf{x}_{k}(t),\mathbf{x}_{-k}(t)) & Otherwise \\
    \end{array}  \right.\] 

\textbf{Update Functions}

{\it Startup Mode Dynamics}: We assume that the system uses a timer to count up to $\tau_I$ seconds. The update function $g_i$ (in Figure~\ref{fig:HIOA_Module}) consists of the two update equations given by 
\begin{align}\label{eq: update_startup_mode}
    \mathbf{p}_{k} [m + 1] &= \mathbf{p}_{k} [m] + \mathbf{p}^{2}_{k} [m] \nonumber \\
    \tau [m+1] &= \tau [m] + h
\end{align}

{\it Normal and Alternate Modes Dynamics}: Here, we assume that the update function depend on the best response to other defenders' investments and previous state. Note that the timer is not used in these modes, thus the corresponding update equation is $\tau [m+1] = 0$. The update function $g$ is given by 
\begin{align}\label{eq: update_normal_mode}
    \mathbf{x}_{k} [m + 1] &= \frac{1}{2} \left(\mathbf{x}_{k} [m-1] + \mathbf{x}^*_{k}[m]  \right) \nonumber \\
    \tau [m+1] &= 0
\end{align}
where $\mathbf{x}^*_{k}[m] \in  argmin_{\mathbf{x}_{k}[m]} \  C_k(\mathbf{x}_{k}[m],\mathbf{x}_{-k}[m])$.

{\it Fail Mode Dynamics}:
In this mode, we assume that the system goes into failure where the probability of successful attack goes to one. Again, note that the timer is not used in this mode, thus the corresponding update equation is $\tau [m+1] = 0$.
The update function $g_o$ in fail mode is given by 
\begin{align}\label{eq: update_fail_mode}
    \mathbf{p}_{k} [m + 1] &= \mathbf{1} \nonumber \\
    \tau [m+1] &= 0
\end{align}

Now, we introduce the parallel decomposition result that enables us composing subnetworks of different defenders to represent the whole large-scale CPS.

\subsection{Parallel Decomposition}

\begin{lemma}\label{lemma: compatability_HIOA}
Given two HIOA $A_1$ and $A_2$, where $A_i$ is defined as the tuple  $(\mathcal{L}_i, \mathcal{X}_i, \mathcal{U}_i, \mathcal{M}_i, \mathcal{G}_i, \mathcal{R}_i, \Delta_i, \mathcal{T}_i, \mathcal{Y}_i, \mathcal{I}_i)$ for $i \in \{1, 2\}$, we say that $A_1$ and $A_2$ are compatible if $\mathcal{X}_1 \cap \mathcal{X}_2 = \phi$, $\mathcal{Y}_1 \cap \mathcal{Y}_2 = \phi$, $\mathcal{Y}_1 \subseteq \mathcal{U}_2$, and $\mathcal{Y}_2 \subseteq \mathcal{U}_1$.
\end{lemma}

\begin{remark}
The parallel composition operation allows compatible HIOA representing two modules (where each module represent a subnetwork of the CPS) to be composed to form a composite module. Note that the compatibility conditions are satisfied with our choice of output and input variables of each HIOA module.
\end{remark}

\subsection{Computing security decisions under combined modelling}

Note that the existing literature  does not effectively capture the significantly more complex scenarios and models that we are considering as part of this paper, involving a mix of static and dynamical nodes. Thus, filling this critical gap is essential for future work. One particular approach that can be pursued is to leverage simulation based optimization (SO) techniques into a broader optimization framework for computing optimal security deployments. Such SO techniques have been widely applied for optimizing complex systems (including building systems and traffic networks)~\cite{osorio2013simulation}, but their use in the broader context of security policies for interdependent CPS is lacking in the literature. SO techniques involve iteratively tuning the optimization parameters based on evaluations of the objective function through a simulator, but face challenges due to the difficulty of evaluating gradients, and in the time taken to run each simulation. Such challenges can be tackled via the use of approximations to the objective functions (here, $f(\cdot)$ and $C_k(\cdot)$) (learned via regression)~\cite{conn2009global}, and by switching between multiple simulators at different levels of resolution, depending on the operating points that are being evaluated~\cite{osorio2013simulation}. Note that creating a systematic approach to integrate such techniques into an optimization framework for computing security deployments for complex CPS through a combination of continuous and combinatorial optimization techniques would be an avenue for future work and beyond the scope of combined modeling of CPS that we consider here.

\section{Real-world Example of combined modeling}\label{sec: Example}

\begin{figure}[t]
\centering
  \includegraphics[width=0.75\linewidth]{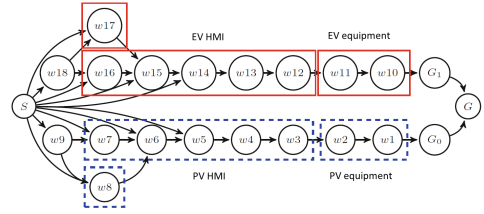}
  \caption{Attack graph of a DER.1 failure scenario adapted from
  \cite{hota2016optimal}. It shows stepping-stone attack steps that can lead to the compromise of PV (i.e., $G_{0}$) or EV (i.e., $G_{1}$). There are two defenders whose critical assets are $G_0$ and $G_1$, while $G$ is a shared critical asset.}
  \label{fig:nescor-attack-graph}
\end{figure}

In this section, we use our proposed combined HIOA model to model a real-world CPS to validate our combined model idea and show the flexibility of such idea in modeling large-scale CPS. We first describe the real-world CPS and then we show how to model such CPS. 

\subsection{DER.1 system description:}\label{subsec: DER_explain}
 
The US National Electric Sector Cybersecurity Organization Resource (NESCOR) Technical Working Group has proposed a framework for evaluating the risks of cyber attacks on the electric grid~\cite{lee2013electric}. 
A distributed energy resource (DER) is described as a cyber-physical system consisting of entities such as generators, storage devices, and electric vehicles, that are part of the energy distribution system \cite{lee2013electric}. The DER.1 failure scenario has been identified as the riskiest failure scenario affecting distributed energy resources according to the NESCOR ranking. Here, there are two critical equipment assets: a PhotoVoltaic (PV) generator and an electric vehicle (EV) charging station. Each piece of equipment is accompanied by a Human Machine Interface (HMI), the only gateway through which the equipment can be controlled. The DER.1 failure scenario is triggered when the attacker gets access to the HMI. The vulnerability of the system may arise due to various reasons, such as hacking of the HMI, or an insider attack. Once the attacker gets access to the system, she changes the DER settings and gets physical access to the DER equipment so that they continue to provide power even during a power system fault.
Through this manipulation, the attacker can cause physical damage to the system, and can even lead to the electrocution of a utility field crew member. 

To analyze the above system within our HIOA model, we follow the model proposed by \cite{hota2016optimal}, which  maps the above high level system overview into an attack graph as shown in Figure \ref{fig:nescor-attack-graph}. In this attack graph, node labels starting with ``$w$'' are used to denote the non-critical assets/equipment used as part of the attack steps, and $ G_{0} $, $G_{1}$, and $ G $ represent the critical assets which are the attacker's goals. 
For the attacker's goals, $ G_{0} $ represents a physical failure of the PV system,  $ G_{1} $ represents a physical failure of the EV system, and $ G $ means that a failure of either type has occurred. The goal $ G $ may signify non-physical losses (e.g., reputation losses) for the DER operator as a result of a successful compromise.
The first defender is responsible for defending the critical asset $ G_{0} $, the second defender for defending $G_{1} $.  
Both defenders share the common asset $ G $. 

\subsection{Combined HIOA Modeling of DER.1}
\begin{figure}[t]
\centering
  \includegraphics[width=0.6\linewidth]{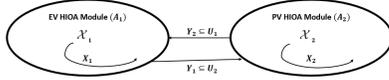}
  \caption{An Example HIOA module for the real-world CPS system (DER.1) composed of two subnetworks. Each subnetwork is represented by a HIOA module.}
  \label{fig:HIOA_Module_DER}
\vspace{-3mm}
\end{figure}

Here, we show the modelling of DER.1 using HIOA. Figure~\ref{fig:HIOA_Module_DER} shows such modeling example where each CPS physical component and its HMI can be represented by a HIOA module (described in Section~\ref{sec: HIOA_classes}). Note that the dynamics and the transitions of each subnetwork are encapsulated in its HIOA combined model. Moreover, we reemphasize that the compatibility condition (in Lemma~\ref{lemma: compatability_HIOA}) is satisfied since the output variables of each module (i.e., the investment of the defender of the corresponding subnetwork) is the input to the other combined HIOA module.

We now present the variables of each HIOA module.
\begin{itemize}
    \item The set of state variables of EV subnetwork module is $\mathcal{X}_1$ = $\{\mathbf{x}_1, \tau, \mathbf{p}_1^0\}$. On the other hand, the set of state variables of PV subnetwork module is $\mathcal{X}_2$ = $\{\mathbf{x}_2, \tau, \mathbf{p}_2^0\}$

\item The set of output variables for EV module is \{$\mathbf{p}_1$, $\mathbf{x}_{1}$\}. The set of output variables for PV module is \{$\mathbf{p}_2$, $\mathbf{x}_{2}$\}.

\item The set of input variables of EV module $\mathcal{U}_1 =  \{Attack\_Risk, Fail\_Event, \mathbf{x}_{-1} = \mathbf{x}_{2}\}$ and similarly $\mathcal{U}_2 =  \{Attack\_Risk, Fail\_Event, \mathbf{x}_{-2} = \mathbf{x}_{1}\}$.
\end{itemize}

Moreover, note that the update functions are calculated using \Crefrange{eq: update_startup_mode}{eq: update_fail_mode} and the trajectories by Equation~\eqref{eq: val_func_1}.

\begin{remark}
We emphasize that our modeling can effectively model any interdependent CPS that can be modeled by attack graphs (e.g., SCADA~\cite{hota2016optimal}, IEEE 300 BUS~\cite{abdallah2020morshed}, and VOIP~\cite{abdallah2020morshed}). However, we omit the details in the interest of the space.
\end{remark}

\section{Conclusion and Future Work}\label{sec: conclusion}
This paper presented a framework that accounts for multiple defenders in a CPS (represented by an attack graph) where the defenders places their investments to protect the target assets. Specifically, we combined HIOA and Game theory for such modeling. We first established the objective function of each defender. We then provided the HIOA model; in particular, we modeled the continuous time nature of the CPS and the transitions between different system states. We validated our model using real-world CPS of a smart energy system. We emphasize that such combination of HIOA and game theory can be applied to model large-scale systems. A future avenue of research would be using simulation based techniques for computing security deployments for complex CPS aided by our combined modeling scheme.

\bibliographystyle{IEEEtran}
\bibliography{refs}

\end{document}